# A multi-plane augmented reality head-up display system based on volume holographic optical elements with large area

### ZHENLV LV, JUAN LIU *, LIANGFA XU

*Beijing Engineering Research Center for Mixed Reality and Advanced Display, School of Optics and Photonics, Beijing Institute of Technology, Beijing, 100081, China*
*\* juanliu@bit.edu.cn*

**Abstract:** The traditional head-up display (HUD) system has the disadvantages of a small area and a single display plane, here we propose and design an augmented reality (AR) HUD system with multi-plane, large area, high diffraction efficiency and a single picture generation unit (PGU) based on holographic optical elements (HOEs). Since volume HOEs have excellent angle selectivity and wavelength selectivity, HOEs of different wavelengths can be designed to display images in different planes. Experimental and simulated results verify the feasibility of this method. Experimental results show that the diffraction efficiencies of the red, green and blue HOEs are 75.2%, 73.1% and 67.5%. And the size of HOEs is 20cm×15cm. Moreover, the three HOEs of red, green and blue display images at different depths of 150cm, 500cm and 1000cm, respectively. In addition, the field of view (FOV) and eye-box (EB) of the system are 12°×10° and 9.5cm×11.2cm. Furthermore, the light transmittance of the system has reached 60%. It is believed that this technique can be applied to the augmented reality navigation display of vehicles and aviation.

## 1. Introduction

The vehicle head-up display (HUD) system can display driving information such as vehicle speed, fuel consumption, warning signals and navigation arrows to the driver through the windshield or combiner [1–4]. With HUD, the driver always looks straight ahead on the road without frequently moving the sight to see the dashboard, thus improving driving safety [5].

The traditional HUD provides vehicle information on a certain depth plane through a small area combiner, without fusion and interaction with the external real environment. However, an augmented reality (AR) HUD can provide richer information to the driver at different depth planes, such as navigation information at a long distance, warning signals at a medium distance, and basic information at a short distance [6–9]. Compared with traditional HUD, AR HUD can greatly improve the driving experience and safety by providing interactive information, which is expected to have a huge automotive and aviation application market.

The research objective of AR HUD is to provide multiple virtual images with different projection distances for the driver. The technical solutions to realize multi-plane display mainly include geometric optics [10–14], dynamic varifocal components [15–19], digital holography [20–24] and volume holography [25]. A HUD with multiple picture generation units (PGUs) and multiple combiners using geometric optics is expensive and bulky, and it is difficult to integrate in the limited space below the dashboard [12,13]. The structure of the HUD with one PGU and one combiner is relatively compact, but the use of free-form surface mirror increases the difficulty of system design and processing [14]. Using dynamic varifocal components, such as liquid lenses [15–17], Pancharatnam–Berry lenses [18], and multilayer polymer switchable liquid crystals [19], is another potential solution for AR HUD, which has been discussed a lot in head-mounted display (HMD). However, compared with HMD, due to the need for a larger eye-box (EB), the size of the optical components in the HUD is much larger. Therefore, high quality, large size and high varifocal speed are the main limiting factors for the application of dynamic components to HUD. Digital holography can realize multi-plane display by calculating and loading holograms of different depths, but it has the problems of low resolution, slow calculation speed and difficulty in color display [24]. In addition, volume holography can also be used for multi-depth plane display. The volume holographic optical elements (HOEs) have great wavelength selectivity and angle selectivity [26,27], which can avoid the crosstalk between the reflected image and the diffracted image. A dual-focal-plane volume holographic HUD that combines a wavelength multiplexing HOE with two PGUs is proposed [25]. The prototype is relatively bulky due to the use of two PGUs. Moreover, the display image is difficult to distinguish in sunlight because of low diffraction efficiency, and there are problems with image distortion and color shift.

In this paper, we propose an AR HUD system with multi-plane, large area, high diffraction efficiency, and a single PGU based on the wavelength selectivity, angle selectivity and imaging function of volume HOEs.

## 2. Basic Idea and principle

The basic structure diagram of the multi-plane AR HUD system is shown in the Fig. 1. The system is mainly composed of three parts: laser projection module, diffuser and holographic combiner. With the characteristics of miniaturization, high brightness and clear imaging at any distance, the laser projection module including Micro-Electro-Mechanical System (MEMS) mirror, optical engine and controller is used to generate the target image. The diffuser receives the image sent by the projection module and is used as the object plane for HOE imaging. The holographic combiner composed of a glass substrate and HOEs with three different wavelengths of red, green and blue is used as the imaging element. In order to achieve multi-plane display, HOEs of different wavelengths correspond to different imaging depths.

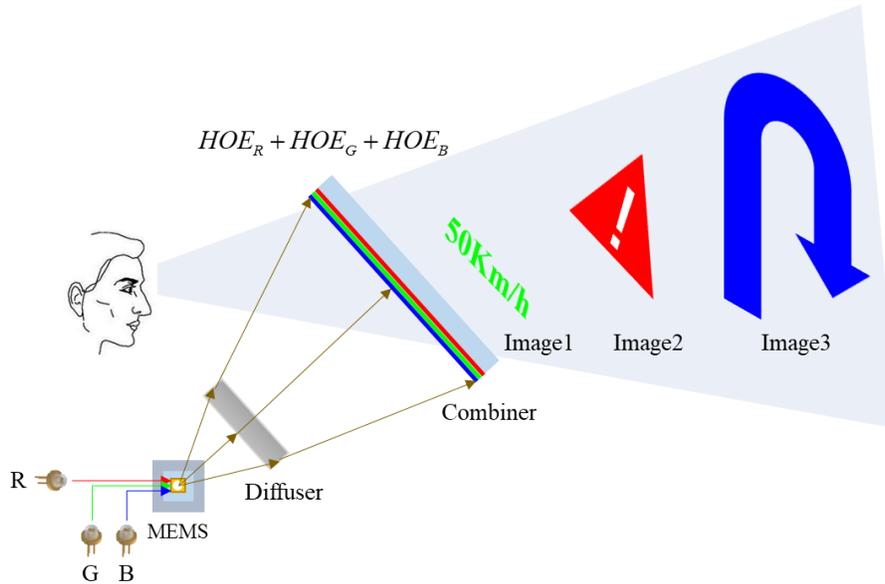

Fig. 1. Schematic diagram of the proposed multi-plane holographic head-up display system.

## 3. Parameter design and numerical simulation

### 3.1 Parameter design

It is necessary to analyze the relationship between objects and images when the HUD system proposed in this paper is used as an imaging system. When analyzing the imaging characteristics of HOEs, the diffuser screen is equivalent to the object plane. The object-image relationship of the system is shown in Fig. 2.

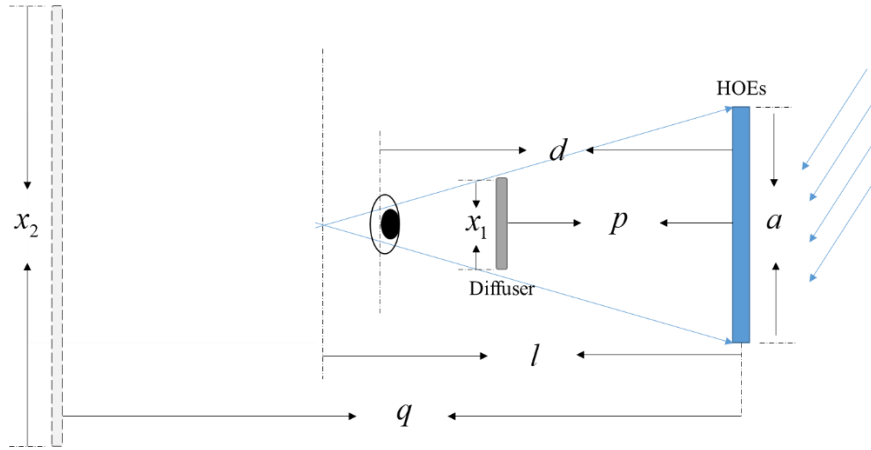

Fig. 2. Diagram of the relationship between object and virtual image.

The image displayed by the HUD system is an enlarged virtual image, so there is the following relationship between the object and the image:

$$\frac{1}{p} - \frac{1}{q} = \frac{1}{f} \tag{1}$$

Where the $p$, $q$ and $f$ are the object distance, image distance and focal length of the system, respectively.

The magnification M between the virtual image and the real image is as follows:

$$M = q/p \tag{2}$$

So the virtual image size $x_2$ is:

$$x_2 = M \times x_1 \tag{3}$$

Where $x_1$ is the size of the image projected on the diffuser.

When used as an imaging element, the reflective volume HOEs can achieve a transmissive display of different planes through the combination of the virtual image imaging function and the reflective function. The virtual image distance actually viewed is the distance between the human eye and the HOEs plus the virtual image distance of the system. Therefore, the field of view (FOV) of the system is as follows:

$$FOV = 2\arctan\left[\frac{x_2}{2(q+d)}\right] \tag{4}$$

Where d is the distance between the human eye and the HOEs.

EB is one of the key indicators of the HUD system. HUD requires a larger eye movement range than HMD. According to the relationship between virtual image, HOEs and human eye, as shown in Fig. 3, the EB of the system can be expressed as:

$$EB = a - (x_2 - a) \times d/q \tag{5}$$

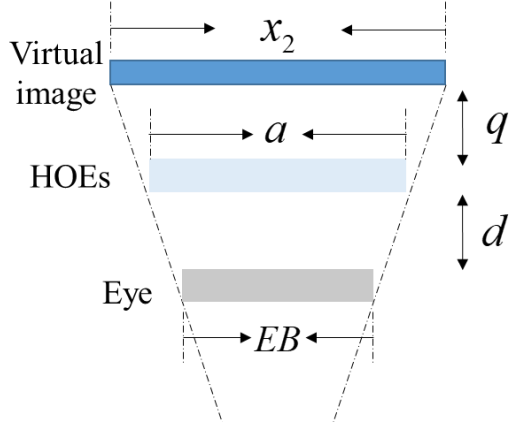

Fig. 3. The relationship between virtual image, HOEs and human eye.

The EB and the FOV have the following relationship:

$$EB = a - 2\tan(FOV/2) \times d \qquad (6)$$

It can be seen from the above formula that when the size of HOEs is constant, there is a negative correlation between FOV and EB. Therefore, the way to ensure that the system has proper FOV and EB at the same time is to increase the size of HOEs, but it will increase the difficulty of design and production.

The recording and reconstruction principles of the multi-plane AR HUD system are shown in Fig. 4. The $O_R$, $O_G$ and $O_B$ are the convergence points of the object light waves corresponding to three different wavelengths modulated by spatial light modulator (SLM). $f_1$, $f_2$ and $f_3$ are the equivalent focal lengths of the HOEs corresponding to the three wavelengths of red, green and blue. $q_1$, $q_2$ and $q_3$ are the virtual image distances of the red, green and blue HOEs at the same object distance.

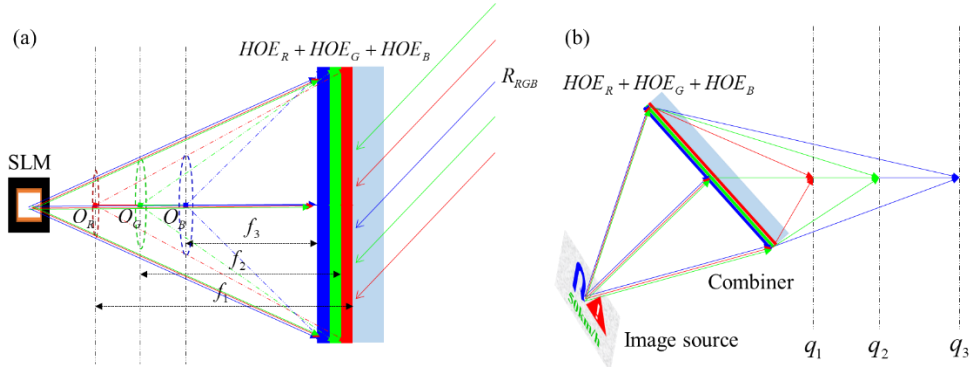

Fig. 4. (a) The recording principle of the system proposed in this paper. (b) The reconstruction principle of the system proposed in this paper.

During recording, the three-wavelength object light waves modulated by the SLM are equivalent to spherical light waves with different degrees of convergence, as shown by the dotted line in the Fig. 4(a). The reference light waves are all parallel light waves. Therefore, the red, green, and blue HOEs as imaging elements have different diopters. During reconstruction, the distances between the target points on the diffuser and the HOEs are constant as shown in the Fig. 4(b), that is, the object distances of the three-wavelength HOEs are the same. For the entire system, the HOEs of the three wavelengths have the same object distances and different focal lengths, thus multi-plane displays with different image distances can be realized.

The design parameters of HOE$_R$, HOE$_G$ and HOE$_B$ are shown in Table 1. The distance between the diffuser and the HOEs is constant at 30 cm. The focal lengths of HOE$_R$, HOE$_G$ and HOE$_B$ are 37.5cm, 31.91cm and 30.93cm respectively. The virtual image distances of HOE$_R$, HOE$_G$ and HOE$_B$ are 150cm, 500cm and 1000cm respectively. The sizes of the three HOEs are all 20cm×15cm. The distance between the human eye and the HOEs is 50cm. In addition, the size of the diffuser is 10cm×10cm. The angle between the reference light and the normal direction of the holographic recording plane is 45°.

Table 1. The design parameters of HOE$_R$, HOE$_G$ and HOE$_B$

|  | HOE$_R$ | HOE$_G$ | HOE$_B$ |
| --- | --- | --- | --- |
| p(cm) | 30 | 30 | 30 |
| q(cm) | 150 | 500 | 1000 |
| f(cm) | 37.5 | 31.91 | 30.93 |
| Size(cm×cm) | 20×15 | 20×15 | 20×15 |
| d(cm) | 50 | 50 | 50 |

*3.2 Numerical simulation*

Based on the designed parameters, we carry out numerical simulation to verify the method proposed in this paper. The simulation process consists of three steps: The first step is the diffraction in the free space between the diffuser and the HOEs. The second step is to superimpose the phase factor of the HOEs. The third step is the diffraction of the free space between the HOEs and the virtual image plane.

The calculation of diffraction in free space is based on Kirchhoff's diffraction integral formula, as shown below:

$$E(x, y) = K \iint A(\xi, \eta) \frac{\exp(jkr)}{r} d\xi d\eta \qquad (7)$$

Where $\xi$ and $\eta$ are the coordinates of the diffractive aperture plane, $K$ is the complex constant, $k$ is the wave vector, and $r$ is the distance from the target point to the center of the spherical wave.

In addition, since the reference light is incident obliquely, the phase factor of the HOEs is as follows:

$$\varphi = -k\sqrt{x^2 + y^2 + r_0^2} + kx\sin\theta \qquad (8)$$

Where $x$ and $y$ are the coordinates of the HOE plane, $r_0$ is the distance from the convergence point of the spherical wave to the HOE, and $\theta$ is the angle of the reference beam deviating from the axis.

The numerical simulation results are shown in Fig. 5. The input image is the three colored letters containing R, G and B loaded on the diffuser. When the observation plane is 150cm away from HOEs, the letter R is clear and the letters G and B are blurred. When the observation plane is 500cm away from HOEs, the letter G is clear and the letters R and B are blurred. When the observation plane is 1000cm away from HOEs, the letter B is clear and the letters R and G are blurred.

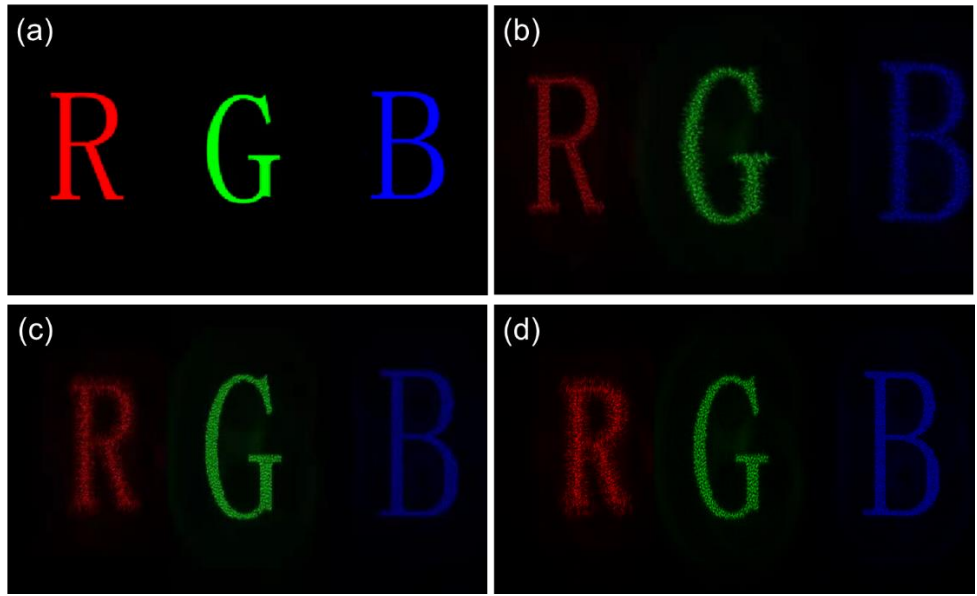

Fig. 5. (a) Input image. (b) Focus on the letter R at 150cm. (c) Focus on the letter G at 500cm. (d) Focus on the letter B at 1000cm.

Therefore, the simulation results prove the feasibility of the method in this paper. However, it can be seen from the simulation results that the displayed image is extended in the vertical direction and the image quality is degraded compared to the input image. The reason for this phenomenon is the astigmatism introduced by the tilt angle between the reference light and the optical axis during recording. The methods of astigmatism correction for the HOEs include adding a cylindrical mirror to the recording optical path [28] and fabricating HOEs on a curved glass substrate [29].

## 4. Experimental verification

Based on the theoretically designed parameters, we record the HOEs corresponding to the three depth planes by the holographic interference exposure method. In order to achieve higher diffraction efficiency, $HOE_R$, $HOE_G$ and $HOE_B$ are exposed independently and stacked on a same glass substrate in layers.

The prototype of the multi-plane AR HUD system consists of three parts: a laser projector, a diffuser and three-layer HOEs, as shown in the Fig. 6(a). The light source is a miniature laser projector manufactured by Ultimems, with a resolution of 1280×760, a brightness of 30 lm, and a horizontal divergence angle of 43.15°. The diffuser is DG100×100-1500 from Thorlabs. In order to achieve lightness and high transmittance, three layers of monochromatic holographic films are stacked on the same glass substrate with a thickness of 2 mm using a vacuum bonding method. The actual produced three-layer HOEs with a size of 20cm×15cm are shown in the Fig. 6(b).

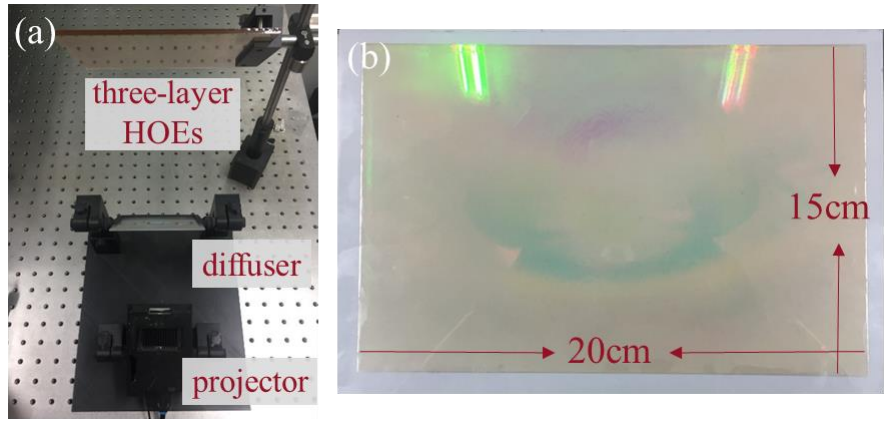

Fig. 6. (a) The prototype of the multi-depth holographic HUD system. (b) the produced three-layer HOEs with a size of 20cm×15cm.

The multi-plane display effect of the prototype has been verified by experiments. Three letters of the same size including R, G, and B are the original input image. The display effects observed by the camera 50cm away from the HOEs are shown in the Fig. 7. When the camera focuses on the letter R at 1.5m, the other two letters G and B are blurred. When the camera focuses on the letter G at 5m, the other two letters R and B are blurred. When the camera focuses on the letter B at 10m, the other two letters R and G are blurred. In addition, the magnifications of the three letters are different due to the different imaging distances. The letter R and the letter B correspond to the minimum and maximum magnification respectively.

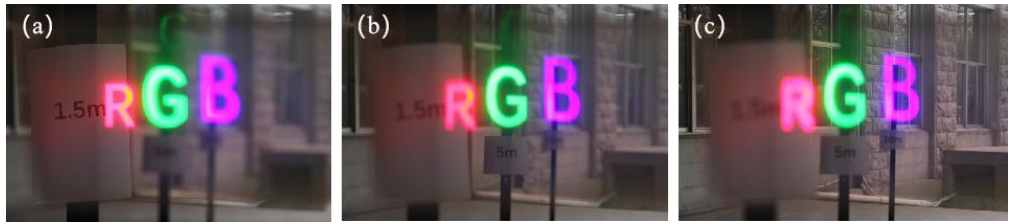

Fig. 7. (a), (b) and (c) are the display effects when the camera focuses on the three letters R, G and B at different distances.

The AR display effect of the HUD system in car navigation is shown in the Fig. 8. The near field of vision displays basic information such as vehicle speed and time through the green HOE. The far field of vision displays the navigation information of road fusion through the blue HOE. In addition, auxiliary information such as warning signs can also be displayed through the red HOE. The maximum FOV of the system is 18°×13.5°, but the EB is only 4.1cm×8.1cm due to the restriction between the FOV and the EB shown in Eq. (6). In order to balance the FOV and the EB, we reduce the size of the loaded image when the size of the HOE is constant. The final FOV and EB of the system are 12°×10° and 9.5cm×11.2cm, respectively. It is worth mentioning that a larger size HOE can be produced to further expand the EB.

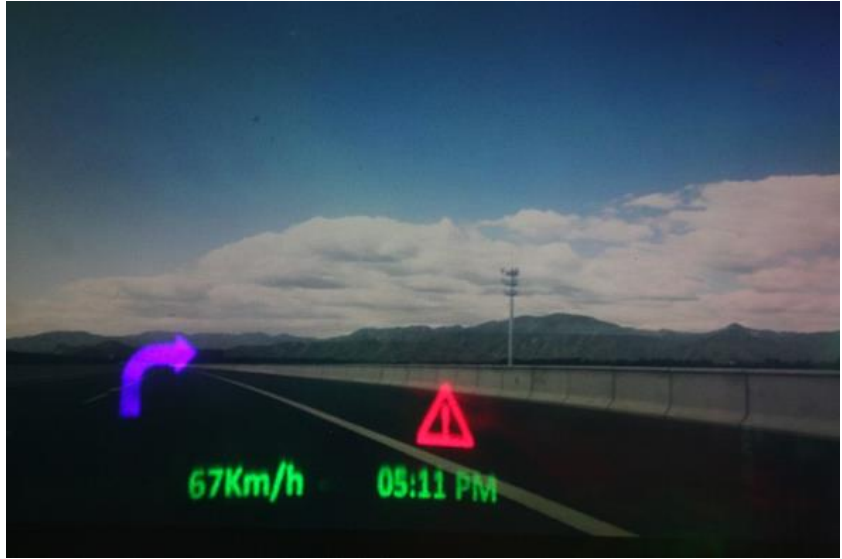

Fig. 8. The AR display effect of the multi-plane holographic HUD system.

## 5. Analysis of results

### 5.1 Diffraction efficiency

For the AR HUD display, too dark navigation signals will be inconvenient for the driver to observe, and too bright signals will affect the driver's judgment of the actual road conditions. Therefore, we analyze the diffraction efficiencies of different gratings. The three wavelengths of the laser used to record the HOEs are 639nm, 532nm and 457nm. The three peak wavelengths of the laser projector used for display measured by the spectrometer are 636nm, 528nm and 449nm. The data are shown in Table 2. Since the recording wavelength and the reconstruction wavelength are not the same, the Bragg mismatch will cause the diffraction efficiency to decrease. Moreover, the greater the wavelength shift, the greater the loss of diffraction efficiency. Therefore, the diffraction efficiency of the three-wavelength HOEs decreases in order of red, green and blue.

Table 2. Recording wavelength and reconstruction wavelength of $HOE_R$, $HOE_G$ and $HOE_B$

|  | recording | reconstruction |
|---|---|---|
| Red | 639nm | 636nm |
| Green | 532nm | 528nm |
| Blue | 457nm | 449nm |

The diffraction efficiencies of the three-wavelength HOEs are measured during reconstruction by loading a pure color image and sampling points to take the average of the efficiency, as shown in the Fig. 9. The diffraction efficiencies of the red, green and blue HOEs measured in the experiment are 75.2%, 73.1% and 67.5%.

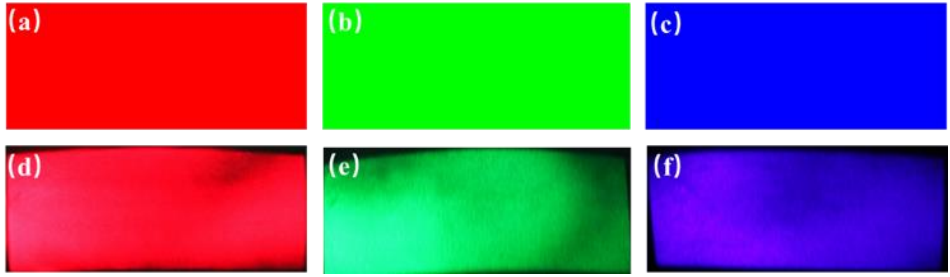

Fig. 9. (a)-(c) Loaded image of red, green and blue HOEs. (d)-(f) Displayed image of red, green and blue HOEs.

A solution to this problem is to use angular offset to compensate for wavelength offset to ensure that Bragg matching conditions are still met when the reconstructed wavelength is inconsistent with the recorded wavelength [30,31]. The relationship between the Bragg mismatch parameter $\delta$ and the angular offset $\Delta\theta$ and the wavelength offset $\Delta\lambda$ is as follows:

$$\delta = \Delta\theta k \sin(\phi-\theta_0) - \Delta\lambda k^2 / 4\pi n_0 \tag{9}$$

Where $k$ is the wave vector, $\theta_0$ is the incident angle of the reference light during recording, $\phi$ is the tilt angle of grating vector, $n_0$ is the bulk index of refraction of the material.

It can be seen from the above formula that the Bragg mismatch parameter can be guaranteed to be zero by simultaneously changing the values of the wavelength offset and the angle offset. In addition, when the diffraction efficiency of the grating is fixed, the intensity of the signal light can be changed by adjusting the brightness of the laser projector.

*5.2 Transmittance*

We also analyze the transmittance of the system, because the transmittance of the HUD is a key parameter in the field of navigation display. With the increase of exposure, the holographic material gradually changes from dark purple to lighter color. Since the exposure dosage during the recording of the HOEs does not reach the maximum recording dosage of the holographic material, the transmittance of the HOEs after recording will increase under sunlight. Limited by the characteristics of the material, the transmittance of the holographic material does not increase indefinitely but gradually tends to a certain value. The light transmittance data under different conditions are shown in Table 3. The transmittance of the glass substrate and the recorded single-layer holographic film are 90% and 88% measured by a light transmittance meter. Experiments show that the transmittance of the HUD system has increased from 56% to 60% after being placed in the sun for a month.

Table 3. Experimental data of light transmittance under different conditions

|  | Glass substrate | Single layer HOE | Glass substrate + Three-layer HOEs | Glass substrate + Three-layer HOEs (one month later) |
|---|---|---|---|---|
| Transmittance | 90% | 88% | 56% | 60% |

## 6. Conclusion

We propose an AR HUD system with multi-plane, large area, high diffraction efficiency and a single PGU. Based on the angular selectivity and wavelength selectivity of the volume HOEs,

HOEs of different wavelengths display images of different depths. We analyze and design the entire system theoretically, and perform numerical simulation and experimental verification. The integrated prototype is composed of a miniature laser projector, a scattering screen and three-layer HOEs. The red, green and blue HOEs are stacked on the same glass substrate with a thickness of 2mm by vacuum bonding. Therefore, the diffraction efficiency of the system is maximized while ensuring lightness and thinness. The diffraction efficiencies of the red, green and blue HOEs measured in the experiment are 75.2%, 73.1% and 67.5%. The HOEs of red, green and blue respectively display images with depths of 150cm, 500cm and 1000cm. The FOV and EB of the system are 12°×10° and 9.5cm×11.2cm. And the size of HOEs is 20cm×15cm. In addition, the transmittance of the system has reached 60%. It is expected that the proposed method will be widely used in augmented reality navigation display, such as vehicles, aviation, etc.